# ON THE BOUNDARY INTENSITIES IN A PLANE PARALLEL SLAB WITH LINEARLY VARYING REFRACTIVE INDEX

Vital Le Dez and Hamou Sadat

Institut PPRIMME, UPR CNRS 3346, Université de Poitiers, 40 Avenue du Recteur Pineau, 86022 Poitiers, France

**Abstract:** An exact expression of the outgoing boundary intensities in the case of a gray semi-transparent medium confined in a plane parallel slab with a refractive index linearly depending on the position is proposed. It is shown that directly using the diffuse reflection law gives in a much easier way exactly the same result as the one obtained with the pseudo-source method.

**Nomenclature**

| | |
|---|---|
| $I_S(\vec{\Omega})$ | intensity leaving a boundary surface for a given direction $\vec{\Omega}$ (Wm$^{-2}$Sr$^{-1}$) |
| $n(z)$ | refractive index |
| $\bar{n}(z)$ | reduced refractive index |
| $E_n$ | exponential integral of order $n$ |
| $a_k$ | reduced refractive index of the lower limit of the cell labelled $k$ |
| $b_k$ | reduced refractive index of the upper limit of the cell labelled $k$ |
| $N$ | cells number |
| $T$ | temperature (K) |

*Greek letters*

| | |
|---|---|
| $\Delta z$ | characteristic cell length |
| $\kappa$ | absorption coefficient (m$^{-1}$) |
| $\varepsilon$ | emissivity of the surfaces |
| $\rho$ | reflection factor of the surfaces |
| $\tau_0$ | optical depth |
| $\bar{\tau}_0$ | generalized optical depth |
| $\sigma$ | Stephan-Boltzmann constant (5.67 10$^{-8}$ Wm$^{-2}$K$^{-4}$) |
| $\vec{\Omega}$ | unit vector of radiation propagation |

*Subscripts*

| | |
|---|---|
| $i, s$ | lower and upper walls respectively |

# I - INTRODUCTION

When the refractive index of a semi-transparent medium (STM) varies continuously, the photons travelling inside the medium follow curved paths governed by the Fermat's principle in the approximation of the geometric optics, which causes important structural modifications of the radiative field inside the bulk and singularly modifies the associated temperature and radiative flux fields. Abundantly examined for the last two decades, the radiative transfer inside such heterogeneous media in the sense of the refractive index, remains nowadays an attractive subject of work and study: first established by Preisendorfer [1] from the transmittivity of a column of STM with a varying index, and revisited by Pomraning [2] who examined the energetic invariant on a given photon path, the determination of the radiative transfer equation (RTE) in such media has produced different and sometimes contradictory works [3-5], although a consensus appears to emerge on the useful form of this equation. For a given RTE form in one-dimensional geometries, Ben Abdallah et al. [6-7] have determined from analytical developments the temperature and radiative flux fields in STM enclosed in parallel slabs bounded by black surfaces at specified temperatures at radiative equilibrium. In one-dimensional geometries such as parallel slabs, Zhu [8] recently extended the curved ray-tracing technique to anisotropically scattering media, while Huang et al. [9, 10] developed a pseudo-sources technique for slabs bounded by diffusely reflecting boundaries, and applied it to radiative transfer inside scattering media with linear and sinusoidal indices. The pseudo-sources method, applied for diffusely reflecting surfaces in a plane geometry, may appear however as a non-intuitive technique, although Xia et al. [11] extended curved ray tracing associated to a pseudo-sources method with diffusely reflecting boundary surfaces to take into account anisotropic scattering.

In this note, instead of using the complex pseudo-source technique, we will simply apply the general law of diffuse reflection on a plane surface, which directly leads to the expressions of the outgoing intensities. Then these expressions are discretized and lead to global coefficients verifying simple closure relations. A comparison with the influencing factors $k_1$ and $k_2$ of the two walls of reference [9] is finally given.

# II - EXPRESSION OF THE INTENSITIES INSIDE THE SLAB

Let us consider a semi-transparent gray absorbing-emitting medium enclosed in a parallel plane slab of depth *d*, bounded by diffusely reflecting surfaces of imposed temperature. The medium is characterised by its absorption coefficient $\kappa$ and its refractive index $n(z)$, supposed linearly variable with the position *z*. The temperature of the lower boundary at $z = 0$ is noted $T_i$ and the refractive index at the same position $n_i$, while for the upper boundary at z = d the temperature and refractive index values are respectively $T_s$ and $n_s$. As in [6], one writes $\bar{n}(z) = \frac{n(z)}{n_i}$, $\tau_0 = \kappa d$ the optical depth and $\bar{\tau}_0 = \frac{\kappa d}{\bar{n}_s - 1}$.

The expression of $I(z, \mu)$ for $\mu \in \,]0,1]$, with $\bar{n}^2(z') > \bar{n}^2(z)$, has been detailed in [6] and is given by :

$$\frac{I(z,\mu)}{n^2(z)} = \frac{I(d,\mu_s)}{n_s^2} e^{-\bar{\tau}_0 \bar{n}(z)\left[\sqrt{\mu^2 + \frac{\bar{n}_s^2 - \bar{n}^2(z)}{\bar{n}^2(z)}} - \mu\right]}$$

$$+ \frac{\kappa\sigma}{\pi\bar{n}(z)} \int_{z'=z}^{d} \frac{\bar{n}(z') T^4(z') e^{-\bar{\tau}_0 \bar{n}(z)\left[\sqrt{\mu^2 + \frac{\bar{n}^2(z') - \bar{n}^2(z)}{\bar{n}^2(z)}} - \mu\right]}}{\sqrt{\mu^2 + \frac{\bar{n}^2(z') - \bar{n}^2(z)}{\bar{n}^2(z)}}} dz' \qquad (6)$$

Similarly the two useful expressions of $I(z,-\mu)$ due to possible internal total reflections are:

* for $\mu \geq \frac{\sqrt{\bar{n}^2(z)-1}}{\bar{n}(z)}$, with $\bar{n}^2(z') < \bar{n}^2(z)$

$$\frac{I(z,-\mu)}{n^2(z)} = \frac{I(0,-\mu_i)}{n_i^2} e^{-\bar{\tau}_0 \bar{n}(z)\left[\mu - \sqrt{\mu^2 - \frac{\bar{n}^2(z)-1}{\bar{n}^2(z)}}\right]}$$

$$+ \frac{\kappa\sigma}{\pi\bar{n}(z)} \int_{z'=0}^{z} \frac{\bar{n}(z') T^4(z') e^{-\bar{\tau}_0 \bar{n}(z)\left[\mu - \sqrt{\mu^2 + \frac{\bar{n}^2(z') - \bar{n}^2(z)}{\bar{n}^2(z)}}\right]}}{\sqrt{\mu^2 + \frac{\bar{n}^2(z') - \bar{n}^2(z)}{\bar{n}^2(z)}}} dz' \qquad (7)$$

* and for $\mu < \frac{\sqrt{\bar{n}^2(z)-1}}{\bar{n}(z)}$, with $\bar{n}^2(z') < \bar{n}^2(z)$

$$\frac{I(z,-\mu)}{n^2(z)} = \frac{I(d,\mu_s)}{n_s^2} e^{-\bar{\tau}_0 \bar{n}(z)\left[\mu + \sqrt{\mu^2 + \frac{\bar{n}_s^2 - \bar{n}^2(z)}{\bar{n}^2(z)}}\right]}$$

$$+ \frac{\kappa\sigma}{\pi\bar{n}(z)} \left\{ \int_{z'=z''}^{z} \frac{\bar{n}(z') T^4(z') e^{-\bar{\tau}_0 \bar{n}(z)\left[\mu - \sqrt{\mu^2 + \frac{\bar{n}^2(z') - \bar{n}^2(z)}{\bar{n}^2(z)}}\right]}}{\sqrt{\mu^2 + \frac{\bar{n}^2(z') - \bar{n}^2(z)}{\bar{n}^2(z)}}} dz' \right.$$

$$\left. + \int_{z'=z''}^{d} \frac{\bar{n}(z') T^4(z') e^{-\bar{\tau}_0 \bar{n}(z)\left[\mu + \sqrt{\mu^2 + \frac{\bar{n}^2(z') - \bar{n}^2(z)}{\bar{n}^2(z)}}\right]}}{\sqrt{\mu^2 + \frac{\bar{n}^2(z') - \bar{n}^2(z)}{\bar{n}^2(z)}}} dz' \right\} \qquad (8)$$

where $z''(z,\mu) = d - \frac{\bar{\tau}_0}{\kappa}\left[\bar{n}_s - \bar{n}(z)\sqrt{1-\mu^2}\right]$

From now we note $I_i^+ = I(0,-\mu_i)$ and $I_s^+ = I(d,\mu_s)$ the intensities leaving the boundary surfaces assumed diffusely reflecting boundaries.

The incident radiation and the radiative flux are completely determined inside the medium when the boundary intensities are obtained, which is the aim of the next section.

# III – DETERMINATION OF THE INTENSITIES $I_i^+$ AND $I_s^+$ FOR DIFFUSELY REFLECTING SURFACES

When the two walls are diffuse reflectors, the law of diffuse reflection writes on the two boundary surfaces:

$$I^+ = \frac{\varepsilon n^2 \sigma T^4}{\pi} + \frac{\rho}{\pi}\int_{\vec{\Omega}\vec{n}<0} I^-|\vec{\Omega}\vec{n}|d\Omega = \frac{\varepsilon n^2 \sigma T^4}{\pi} + 2\rho\int_{\mu=0}^{1} I^-\mu d\mu \qquad (11)$$

where $I^+$ is the intensity leaving the surface and $I^-$ is the intensity incoming on this same surface. For diffusely reflecting surfaces, the leaving intensity is not depending on the propagation angle. The incident intensity $I_i^-$ on the lower surface is deduced from (6) with $z = 0$ and:

$$\frac{I_i^-}{n_i^2} = \frac{I_s^+}{n_s^2} e^{-\bar{\tau}_0\left(\sqrt{\mu^2+\bar{n}_s^2-1}-\mu\right)} + \frac{\kappa\sigma}{\pi}\int_{z'=0}^{d} \frac{\bar{n}(z')T^4(z')e^{-\bar{\tau}_0\left[\sqrt{\mu^2+\bar{n}^2(z')-1}-\mu\right]}}{\sqrt{\mu^2+\bar{n}^2(z')-1}} dz' \qquad (12)$$

The intensity on the upper boundary for angles verifying $\mu \geq \frac{\sqrt{\bar{n}_s^2-1}}{\bar{n}_s}$ is obtained from (7) at $z = d$, whence:

$$\frac{I_s^-}{n_s^2} = \frac{I_i^+}{n_i^2} e^{-\bar{\tau}_0\bar{n}_s\left(\mu - \sqrt{\mu^2 - \frac{\bar{n}_s^2-1}{\bar{n}_s^2}}\right)} + \frac{\kappa\sigma}{\pi\bar{n}_s}\int_{z'=0}^{d} \frac{\bar{n}(z')T^4(z')e^{-\bar{\tau}_0\bar{n}_s\left[\mu - \sqrt{\mu^2+\frac{\bar{n}^2(z')-\bar{n}_s^2}{\bar{n}_s^2}}\right]}}{\sqrt{\mu^2 + \frac{\bar{n}^2(z')-\bar{n}_s^2}{\bar{n}_s^2}}} dz' \qquad (13)$$

For angles such that $\mu < \frac{\sqrt{\bar{n}_s^2-1}}{\bar{n}_s}$, one uses Eq. (8) yielding to:

$$\frac{I_s^-}{n_s^2} = \frac{I_s^+}{n_s^2} e^{-2\bar{\tau}_0\bar{n}_s\mu}$$

$$+ \frac{2\kappa\sigma}{\pi\bar{n}_s}\int_{z'=d\frac{\bar{n}_s\sqrt{1-\mu^2}-1}{\bar{n}_s-1}}^{d} \frac{\bar{n}(z')T^4(z')e^{-\bar{\tau}_0\bar{n}_s\mu}\cosh\left[\bar{\tau}_0\bar{n}_d\sqrt{\mu^2+\frac{\bar{n}^2(z')-\bar{n}_s^2}{\bar{n}_s^2}}\right]}{\sqrt{\mu^2+\frac{\bar{n}^2(z')-\bar{n}_s^2}{\bar{n}_s^2}}} dz' \qquad (14)$$

Therefore Eq. (11) immediately leads to:

$$\frac{I_i^+}{n_i^2} - 2\rho_i\bar{n}_s^2\Theta_s\frac{I_s^+}{n_s^2} = \frac{\varepsilon_i\sigma T_i^4}{\pi} + \frac{2\rho_i\kappa\sigma}{\pi}\int_{z'=0}^{d} \bar{n}(z')K'_2(z')T^4(z')dz' \qquad (15)$$

The angularly integrated kernels appearing in Eq. (15) are given by:

$$\Theta_s = \frac{E_3(\bar{\tau}_0)}{\bar{n}_s} - \frac{\bar{\tau}_0 E_4(\bar{\tau}_0) + E_5(\bar{\tau}_0) - \left[\bar{\tau}_0\sqrt{\bar{n}_s^2 - 1}E_4\left(\bar{\tau}_0\sqrt{\bar{n}_s^2 - 1}\right) + E_5\left(\bar{\tau}_0\sqrt{\bar{n}_s^2 - 1}\right)\right]}{\bar{\tau}_0^2 \bar{n}_s^2}$$

$$K'_2(z') = E_2(\kappa z') - \frac{E_3(\kappa z') - E_3\left[\bar{\tau}_0\sqrt{\bar{n}^2(z') - 1}\right]}{\bar{\tau}_0}$$

Similarly for the outgoing intensity from the upper boundary, the same application of the reflection law yields to:

$$(1 - 2\rho_s\Theta_i)\frac{I_s^+}{n_s^2} - 2\rho_s\Theta_s\frac{I_i^+}{n_i^2} = \frac{\varepsilon_s\sigma T_s^4}{\pi} + \frac{2\rho_s\kappa\sigma}{\pi\bar{n}_s}\left[\int_{z'=0}^{d}\bar{n}(z')K'_1(z')T^4(z')dz' + L'\right] \quad (16)$$

where:

$$K'_1(z') = E_2[\kappa(d - z')] - \frac{\sqrt{\bar{n}_s^2 - 1}}{\bar{n}_s}E_2\left\{\bar{\tau}_0\left[\sqrt{\bar{n}_s^2 - 1} - \sqrt{\bar{n}^2(z') - 1}\right]\right\}$$
$$+ \frac{E_3[\kappa(d - z')] - E_3\left\{\bar{\tau}_0\left[\sqrt{\bar{n}_s^2 - 1} - \sqrt{\bar{n}^2(z') - 1}\right]\right\}}{\bar{\tau}_0\bar{n}_s}$$

$$\Theta_i = \frac{1}{4\bar{\tau}_0^2\bar{n}_s^2}\left[1 - \left(2\bar{\tau}_0\sqrt{\bar{n}_s^2 - 1} + 1\right)e^{-2\bar{\tau}_0\sqrt{\bar{n}_s^2 - 1}}\right]$$

The integral $L'$ being defined by:

$$L' = \frac{2\kappa}{\bar{\tau}_0\bar{n}_s}\int_{z''=0}^{d}\int_{z'=z''}^{d}\frac{\bar{n}(z'')\bar{n}(z')T^4(z')e^{-\bar{\tau}_0\sqrt{\bar{n}_s^2 - \bar{n}^2(z'')}}\cosh\left[\bar{\tau}_0\sqrt{\bar{n}^2(z') - \bar{n}^2(z'')}\right]}{\sqrt{\bar{n}^2(z') - \bar{n}^2(z'')}}dz'\,dz''$$

The two equations (15-16) lead to the two boundary intensities:

$$\frac{\pi}{\sigma}(1 - 2\rho_s\Theta_i - 4\rho_i\rho_s\bar{n}_s^2\Theta_s^2)\frac{I_i^+}{n_i^2} - [(1 - 2\rho_s\Theta_i)\varepsilon_iT_i^4 + 2\varepsilon_s\rho_i\bar{n}_s^2\Theta_sT_s^4]$$
$$= 2\kappa\rho_i\left\{\int_{z'=0}^{d}\bar{n}(z')[2\rho_s\bar{n}_s\Theta_sK'_1(z') + (1 - 2\rho_s\Theta_i)K'_2(z')]T^4(z')dz' \right. \quad (17)$$
$$\left. + 2\rho_s\bar{n}_s\Theta_sL'\right\}$$

$$\frac{\pi}{\sigma}(1 - 2\rho_s\Theta_i - 4\rho_i\rho_s\bar{n}_s^2\Theta_s^2)\frac{I_s^+}{n_s^2} - (2\varepsilon_i\rho_s\Theta_sT_i^4 + \varepsilon_sT_s^4)$$
$$= \frac{2\kappa\rho_s}{\bar{n}_s}\left\{\int_{z'=0}^{d}\bar{n}(z')[K'_1(z') + 2\rho_i\bar{n}_s\Theta_sK'_2(z')]T^4(z')dz' + L'\right\} \quad (18)$$

## IV – DISCRETE FORM OF THE SOLUTION

The slab is divided in isothermal equally spaced cells of depth $\Delta z$, the cell labelled $i$ being centred at the point $z_j = \left(j - \frac{3}{2}\right)\Delta z$ with $2 \leq j \leq N - 1$ where $N$ is the number of cells. The two boundary surfaces themselves are 0 depth cells such that $j = 1$ corresponds to $z_1 = 0$ and $j = N$ corresponds to $z_N = d$. Then $z_2 = z_1 + \frac{\Delta z}{2}$ and $z_{N-1} = z_N - \frac{\Delta z}{2}$ with $\Delta z = \frac{d}{N-2}$.

For the outgoing intensities, one writes for isothermal cells:

$$\kappa \int_{z'=0}^{d} \bar{n}(z')K'_1(z')T^4(z')dz' = \kappa \sum_{k=2}^{N-1} T_k^4 \int_{z'=(k-2)\Delta z}^{(k-1)\Delta z} \bar{n}(z')K'_1(z')dz' = \sum_{k=2}^{N-1} M_k T_k^4 \qquad (19)$$

$$\kappa \int_{z'=0}^{d} \bar{n}(z')K'_2(z')T^4(z')dz' = \sum_{k=2}^{N-1} N_k T_k^4 \qquad (20)$$

$$\kappa L' = \sum_{k=2}^{N-1} P_k T_k^4 \qquad (21)$$

where the coefficients $M_k$, $N_k$ and $P_k$ are fully detailed in the Annex. Writing $R_k = M_k + P_k$ finally leads to the two following relations:

$$\sum_{k=2}^{N-1} N_k = \frac{1}{2} - \bar{n}_s^2 \Theta_s \qquad \sum_{k=2}^{N-1} R_k = \bar{n}_s \left(\frac{1}{2} - \Theta_s - \Theta_i\right)$$

The coefficients $N_k$ and $R_k$ can easily be related to the coefficients $A_k$ and $C_k$ of reference [9] where a pseudo-source method has been used.

Then the two outgoing intensities can be written in the discrete form:

$$\frac{I_i^+}{n_i^2} = \frac{\sigma}{\pi} \sum_{k=1}^{N} S_k T_k^4 \qquad \frac{I_s^+}{n_s^2} = \frac{\sigma}{\pi} \sum_{k=1}^{N} U_k T_k^4 \qquad (22)$$

where the different coefficients $S_k$ and $U_k$, which verify the conservation relations $\sum_{k=2}^{N-1} S_k = 1$ and $\sum_{k=2}^{N-1} U_k = 1$, are detailed in the Annex. Note that the quantity $\Sigma$ defined by $\Sigma = 1 - 2\rho_s\Theta_i - 4\rho_i\rho_s\bar{n}_s^2\Theta_s^2$ is simply the determinant $D = \begin{vmatrix} 1 & -2\rho_i\bar{n}_s^2\Theta_s \\ -2\rho_s\Theta_s & 1 - 2\rho_s\Theta_i \end{vmatrix}$ of the linear system (15-16), directly related to the influencing factors of the two walls $k_1$ and $k_2$ of reference [9], with $k_2 = 2\bar{n}_s^2\Theta_s$ and $1 - K_1\rho_s = \Sigma$

## VII – CONCLUSION

In this paper, we directly obtained the exact expressions of the boundary intensities inside a parallel slab filled with a semi-transparent medium for which the refractive index is linearly varying, from the law of the diffuse reflection without using the pseudo-sources method. These intensities are spatially discretized, yielding to a set of coefficients verifying simple closure relations. The validity of the resulting intensities has been compared to the pseudo-source method, and according to us, the direct application of the diffuse reflection law, which appears to be strictly equivalent to the pseudo-sources method, is of easier use.

## References

[1] R. W. Preisendorfer, Radiative Transfer on Discrete Spaces, *International Series of Monographs on pure and applied Mathematics*, Vol. 74, Pergamon Press, 1965.

[2] G. C. Pomraning, The Equations of Radiation Hydrodynamics, Pergamon Press N. Y, 1973

[3] L. Martí-López and J. Bouza-Domínguez, J.C. Hebden, S. R. Arridge and R.A. Martínez-Celorio, Validity conditions for the radiative transfer equation, J. Opt. Soc. Am. A, Vol. 20, No. 11, pp. 2046-2056, 2003

[4] Jean-Michel Tualle, Link between the laws of geometrical optics and the radiative transfer equation in media with a spatially varying refractive index, Optics Communications 281, 14, pp. 3631-3635, 2008

[5] M. L. Shendeleva and J. A. Molloy, Scaling property of the diffusion equation for light in a turbid medium with varying refractive index, J. Opt. Soc. Am. A, Vol. 24, No. 9, pp. 2902-2910, 2007

[6] Ben Abdallah P. and Le Dez V., Temperature Field inside an Absorbing-Emitting Semi-Transparent Slab at Radiative Equilibrium with Variable Spatial Refractive Index, J. Quant. Spectrosc. Radiat. Transfer 65, pp 595-608, 2000

[7] Ben Abdallah P. and Le Dez V., Radiative flux field inside an absorbing-emitting semi-transparent slab with variable spatial refractive index at radiative conductive coupling, J. Quant. Spectrosc. Radiat. Transfer 67, pp 125-137, 2000

[8] Ke-Yong Zhu, Yong Huang and Jun Wang, Curved ray tracing method for one-dimensional radiative transfer in the linear-anisotropic scattering medium with graded index, J. Quant. Spectrosc. Radiat. Transfer, Vol. 112 pp 377-383, 2011

[9] Huang Y., Xia XL, Tan HP, Temperature field of radiative equilibrium in a semitransparent slab with a linear refractive index and gray walls, J. Quant. Spectrosc. Radiat. Transfer, Vol. 74, pp. 249-261, 2002

[10] Y. Huang, X.L. Xia, H.P. Tan, Radiative intensity solution and thermal emission analysis of a semitransparent medium layer with a sinusoidal refractive index, J. Quant. Spectrosc. Radiat. Transfer, Vol. 74, pp. 217-233, 2002

[11] Xin-Lin Xia, Yong Huang, He-Ping Tan and Xia-Bin Zhang, Simultaneous radiation and conduction heat transfer in a graded index semi-transparent slab with gray boundaries, International Journal of Heat and Mass Transfer, Vol. 45, No 13, pp. 2673-2688, 2002

# ANNEX

Like in [6] we use the following notations at each cell labelled $k$: $\bar{n}[(k-1)\Delta z] = \frac{1}{2}(\bar{n}_{k+1} + \bar{n}_k) = a_k$ and $\bar{n}[(k-2)\Delta z] = b_k = a_{k-1}$.

## Coefficients for the outgoing intensities:

$$M_k = \left[\frac{uE_4(u) + E_5(u)}{\bar{\tau}_0^2 \bar{n}_s} + \left(\frac{u}{\bar{\tau}_0} - \bar{n}_s\right)E_3(u)\right]_{\bar{\tau}_0(\bar{n}_s-a_k)}^{\bar{\tau}_0(\bar{n}_s-b_k)}$$

$$- \left[\frac{uE_4(u) + E_5(u)}{\bar{\tau}_0^2 \bar{n}_s} + \frac{\sqrt{\bar{n}_s^2-1}}{\bar{n}_s}\left(\frac{u}{\bar{\tau}_0} - \sqrt{\bar{n}_s^2-1}\right)E_3(u)\right]_{\bar{\tau}_0\left(\sqrt{\bar{n}_s^2-1}-\sqrt{a_k^2-1}\right)}^{\bar{\tau}_0\left(\sqrt{\bar{n}_s^2-1}-\sqrt{b_k^2-1}\right)}$$

and:

$$N_k = \left[\frac{uE_4(u) + E_5(u)}{\bar{\tau}_0^2} - \left(\frac{u}{\bar{\tau}_0} + 1\right)E_3(u)\right]_{\bar{\tau}_0(b_k-1)}^{\bar{\tau}_0(a_k-1)} - \frac{1}{\bar{\tau}_0^2}[uE_4(u) + E_5(u)]_{\bar{\tau}_0\sqrt{b_k^2-1}}^{\bar{\tau}_0\sqrt{a_k^2-1}}$$

Introducing then the function $H(z',z'')$ such that $H(z',z'') = \dfrac{\bar{n}(z'')\bar{n}(z\prime)e^{-\bar{\tau}_0\sqrt{\bar{n}_s^2-\bar{n}^2(z'')}}\cosh\left[\bar{\tau}_0\sqrt{\bar{n}^2(z\prime)-\bar{n}^2(z'')}\right]}{\sqrt{\bar{n}^2(z\prime)-\bar{n}^2(z'')}}$ allows writing for the last integral $L'$:

$$\kappa L' = \sum_{k=2}^{N-1} P_k T_k^4$$

$$= \frac{2\kappa^2}{\bar{\tau}_0 \bar{n}_s} \sum_{k=2}^{N-1} \int_{z''=(k-2)\Delta z}^{(k-1)\Delta z} \left[T_k^4 \int_{z'=z''}^{(k-1)\Delta z} H(z',z'')dz'\right.$$

$$\left. + \sum_{m=k+1}^{N-1} T_m^4 \int_{z'=(m-2)\Delta z}^{(m-1)\Delta z} H(z',z'')dz'\right] dz''$$

Whence replacing $H$ by its real value finally leads to, when using the variable change $u = \sqrt{\bar{n}_s^2 - \bar{n}^2(z'')}$:

$$P_k = \frac{1}{\bar{n}_s}\langle \int_{u=\sqrt{\bar{n}_s^2-a_k^2}}^{\sqrt{\bar{n}_s^2-1}} u\left\{e^{-\bar{\tau}_0\left[u-\sqrt{u^2-(\bar{n}_s^2-a_k^2)}\right]} - e^{-\bar{\tau}_0\left[u+\sqrt{u^2-(\bar{n}_s^2-a_k^2)}\right]}\right\} du$$

$$- \int_{u=\sqrt{\bar{n}_s^2-b_k^2}}^{\sqrt{\bar{n}_s^2-1}} u\left\{e^{-\bar{\tau}_0\left[u-\sqrt{u^2-(\bar{n}_s^2-b_k^2)}\right]} - e^{-\bar{\tau}_0\left[u+\sqrt{u^2-(\bar{n}_s^2-b_k^2)}\right]}\right\} du\rangle$$

Then noting $Q^{\pm}(\alpha) = \int_{u=\sqrt{\bar{n}_s^2-\alpha^2}}^{\sqrt{\bar{n}_s^2-1}} u e^{-\bar{\tau}_0 \left[u \pm \sqrt{u^2-(\bar{n}_s^2-\alpha^2)}\right]} du$ where $\alpha < \bar{n}_s$, the calculation yields to:

$$Q^-(\alpha) - Q^+(\alpha)$$
$$= \frac{1}{\bar{\tau}_0^2} [E_5(u) + uE_4(u)]_{\bar{\tau}_0\left(\sqrt{\bar{n}_s^2-1}-\sqrt{\alpha^2-1}\right)}^{\bar{\tau}_0\left(\sqrt{\bar{n}_s^2-1}+\sqrt{\alpha^2-1}\right)}$$
$$+ \sqrt{\bar{n}_s^2 - 1}\sqrt{\alpha^2 - 1} \left\{ E_3\left[\bar{\tau}_0\left(\sqrt{\bar{n}_s^2-1} + \sqrt{\alpha^2-1}\right)\right] + E_3\left[\bar{\tau}_0\left(\sqrt{\bar{n}_s^2-1} - \sqrt{\alpha^2-1}\right)\right]\right\}$$

Writing $R_k = M_k + P_k$ finally leads to:

$$R_k = \left[\frac{uE_4(u) + E_5(u)}{\bar{\tau}_0^2 \bar{n}_s} - \left(\bar{n}_s - \frac{u}{\bar{\tau}_0}\right) E_3(u)\right]_{\bar{\tau}_0(\bar{n}_s-a_k)}^{\bar{\tau}_0(\bar{n}_s-b_k)}$$
$$+ \left[\frac{uE_4(u) + E_5(u)}{\bar{\tau}_0^2 \bar{n}_s} + \frac{\sqrt{\bar{n}_s^2-1}}{\bar{n}_s}\left(\frac{u}{\bar{\tau}_0} - \sqrt{\bar{n}_s^2-1}\right) E_3(u)\right]_{\bar{\tau}_0\left(\sqrt{\bar{n}_s^2-1}+\sqrt{b_k^2-1}\right)}^{\bar{\tau}_0\left(\sqrt{\bar{n}_s^2-1}+\sqrt{a_k^2-1}\right)}$$

Noting then $\Sigma = 1 - 2\rho_s \Theta_i - 4\rho_i \rho_s \bar{n}_s^2 \Theta_s^2$, yields for the two outgoing intensities to:

$$S_1 = \frac{\varepsilon_i(1 - 2\rho_s\Theta_i)}{\Sigma} \qquad U_1 = \frac{2\varepsilon_i\rho_s\Theta_s}{\Sigma}$$
$$S_k = \frac{2\rho_i}{\Sigma}[N_k(1 - 2\rho_s\Theta_i) + 2\rho_s\bar{n}_s\Theta_s R_k] \quad 2 \le k \le N-1$$
$$U_k = \frac{2\rho_s}{\bar{n}_s\Sigma}(2\rho_i\bar{n}_s\Theta_s N_k + R_k)$$
$$S_N = \frac{2\varepsilon_s\rho_i\bar{n}_s^2\Theta_s}{\Sigma} \qquad U_N = \frac{\varepsilon_s}{\Sigma}$$